\documentclass[12pt]{article}
\usepackage{amssymb}
\makeatletter
\@addtoreset{equation}{section}
\makeatother

\begin{document}
\begin{center}
{\Large \bf
Indeterministic Quantum Gravity and Cosmology} \\[0.5cm]
{\large\bf VIII. Gravilon: Gravitational Autolocalization}
\\[1.5cm]
{\bf Vladimir S.~MASHKEVICH}\footnote {E-mail:
mashkevich@gluk.apc.org}  \\[1.4cm]
{\it Institute of Physics, National academy
of sciences of Ukraine \\
252028 Kiev, Ukraine} \\[1.4cm]
\vskip 1cm

{\large \bf Abstract}
\end{center}

This paper is a sequel to the series of papers [1-7].
Gravitational autolocalization of a body is considered. A
self-consistent problem is solved: A quantum state of the
center of mass of the body gives rise to a classical
gravitational field, and the state, on the other hand, is an
eigenstate in the field. We call a resulting solution gravilon.
Gravilons are classified, and their properties are studied.
Gravitational autolocalization is predominantly a macroscopic
effect. The motion of a gravilon as a whole is classical.

\newpage

\hspace*{3 cm}
\begin{minipage}[b]{13 cm}
Watching the rolling ball, Mr. Tompkins noticed to his great
surprise that the ball began to 'spread out'. This was the only
expression he could find for the strange behavior of the ball
which, moving across the green field, seemed to become more
and more washed out, losing its sharp contours.
\end{minipage}
\begin{flushright}
George Gamow, Mr. Tompkins in Wonderland \vspace*{0.8 cm}
\end{flushright}

\begin{flushleft}
\hspace*{0.5 cm} {\Large \bf Introduction}
\end{flushleft}

The matter of the universe is governed by quantum
laws. Yet macroscopic bodies exhibit a pronounced classical
behavior. This leads to the problem of classicality (see,
e.g., [8]). The most essential aspect of the problem is the
well-marked localization of macroscopic objects. We quote
Penrose [9]: ''Why, then, do we not experience macroscopic
bodies, say cricket balls, or even people, having two
completely different locations at once? This is a profound
question, and present-day quantum theory does not really
provide us with a satisfying answer''. Indeed, in quantum
mechanics, for a stationary state of a free body, the center
of mass of the body is not localized whatsoever.

We argue that the classicality problem, specifically the
localization problem, cannot be solved unless classical
elements have been incorporated into a theory from the
outset. But it is only spacetime, not matter, that allows a
classical description, so that we argue for semiclassical
gravity and its using for solving the localization problem.

Indeterministic quantum gravity and cosmology  (IQGC)---the
theory being developed in this series of papers---is based
on a variant of semiclassical gravity, so that IQGC provides a
natural groundwork for handling the localization problem.

The idea is to consider a self-consistent problem: A quantum
state of the center of mass of a body gives rise to a classical
gravitational field, and the state, on the other hand,
is an eigenstate in the field. A model solution to the problem
is very simple.

We call a gravitationally autolocalized body
gravilon: gravilon={\bf gravi}ty+
{\bf lo}calizatio{\bf n}. The main results are as follows.

Let $a_{0}$ be the radius of a body with a mass $M$, $r_{0}$
be the radius of the wave function of the center of mass of
the body, $a=a_{0}+r_{0}$, and $m_{P}$ be the Planck mass.
A light (respectively heavy) gravilon is that with
$M\lesssim m_{P}$ (respectively $M\gg m_{P}$); a fuzzy
(respectively
quasiclassical) gravilon is that with $r_{0}\gtrsim a_{0}$
(respectively $r_{0}\ll a_{0}$).

A gravilon is quasiclassical iff $(a_{0}/l_{P})
(M/m_{P})^{3}\gg 10$ where $l_{P}$ is the Planck length.
For the quasiclassical gravilon with a constant density,
$r_{0}\propto 1/a_{0}^{3/2}$.

A heavy gravilon is quasiclassical, a fuzzy one is light.
A light gravilon may be both fuzzy and quasiclassical.

For any gravilon the conditions $a\gg l_{P}(M/m_{P}),\;\;
a\gtrsim l_{P}(m_{P}/M)^{3}$ should be fulfilled; it follows
$a\gg l_{P}$.

Gravitational autolocalization is predominantly a macroscopic
effect.

The motion of a gravilon as a whole is classical.

\section{Model}

Consider a ball with a radius $a_{0}$, a mass $M$, and a
constant density $\rho$, so that
\begin{equation}
M=\frac{4\pi}{3}a_{0}^{3}\rho.
\label{1.1}
\end{equation}
Let $r_{0}$ be the radius of the wave function of the center
of mass of the ball, and
\begin{equation}
a=a_{0}+r_{0}.
\label{1.2}
\end{equation}
We use the Newtonian approximation for gravitational potential
$\Phi(r)$ and put
\begin{equation}
\Phi(r)=\kappa M\left\{
\begin{array}{rcl}
(r^{2}-3a^{2})/2a^{3}\qquad {\rm for}\quad r\le a\\
(-1/r)\qquad {\rm for}\quad r\ge a,
\end{array}
\right.
\label{1.3}
\end{equation}
where $\kappa$ is the gravitational constant.

Now we consider a quantum particle with the mass $M$ in
a potential well
\begin{equation}
U(r)=\frac{\kappa M^{2}}{2a^{3}}r^{2},
\label{1.4}
\end{equation}
i.e., a harmonic oscillator. From the relation
\begin{equation}
U(r)=\frac{M\omega^{2}r^{2}}{2}
\label{1.5}
\end{equation}
it follows for the frequency
\begin{equation}
\omega=\frac{(\kappa M)^{1/2}}{a^{3/2}}.
\label{1.6}
\end{equation}
We have (from here on $\hbar=1, c=1$)
\begin{equation}
r_{0}=\sqrt{\frac{2}{M\omega}}=\frac{\sqrt{2}a^{3/4}}
{\kappa^{1/4}M^{3/4}},
\label{1.7}
\end{equation}
which, in view of eq.(\ref{1.2}), is an equation for
$r_{0}$.

\section{Conditions}

The condition for the Newtonian approximation is
\begin{equation}
2|\Phi(0)|\ll 1.
\label{2.1}
\end{equation}
The condition that a black hole does not form is
\begin{equation}
a>2\kappa M.
\label{2.2}
\end{equation}
The condition that there is no creation of particle-antiparticle
pairs is
\begin{equation}
|\Phi(0)|<\Phi_{\rm critical},\quad \Phi_{\rm critical}>1.
\label{2.3}
\end{equation}
It is seen that the condition (\ref{2.1}) implies the conditions
(\ref{2.2}),(\ref{2.3}). Thus we assume that
\begin{equation}
a\gg l_{P}\frac{M}{m_{P}}
\label{2.4}
\end{equation}
holds where $l_{P}=t_{P}$ is the Planck length and/or time and
$m_{P}$ is the Planck mass ($\kappa=t_{P}^{2}, m_{P}=1/t_{P}$).

Furthermore, the condition
\begin{equation}
\frac{\omega}{2}\ll |\Phi(0)|M
\label{2.5}
\end{equation}
should hold, which results in
\begin{equation}
a\gg \frac{1}{9}l_{P}\left(\frac{m_{P}}{M}\right)^{3}.
\label{2.6}
\end{equation}
Thus we have obtained the conditions
\begin{equation}
a\gg l_{P}\frac{M}{m_{P}},\qquad a\gtrsim l_{P}\left(
\frac{m_{P}}{M} \right)^{3}\equiv a_{1}.
\label{2.7}
\end{equation}
It follows from those
\begin{equation}
a\gg l_{P}.
\label{2.8}
\end{equation}

\section{Solution}

For the sake of simplicity, we drop the factor $\sqrt{2}$ in
eq.(\ref{1.7}) and obtain
\begin{equation}
r_{0}^{4}=l_{P}\left( \frac{m_{P}}{M} \right)^{3}(r_{0}+
a_{0})^{3}.
\label{3.1}
\end{equation}
Let us introduce quantities
\begin{equation}
x=\frac{r_{0}}{a_{0}},\qquad \alpha=\frac{a_{0}}
{a_{1}},
\label{3.2}
\end{equation}
then eq.(\ref{3.1}) takes the form of
\begin{equation}
\frac{(1+x)^{3}}{x^{4}}=\alpha.
\label{3.3}
\end{equation}
We call the quantity $\alpha$ characteristic parameter.

Let
\begin{equation}
{\rm I\:i}.\qquad \qquad \qquad   \alpha\ll 1.
\label{3.4}
\end{equation}
Then
\begin{equation}
x\gg 1,\qquad x\cong \frac{1}{\alpha},
\label{3.5}
\end{equation}
so that
\begin{equation}
r_{0}\cong a_{1}\qquad a_{0}\ll a_{1}\qquad
r_{0}\gg a_{0}\qquad a\cong a_{1}.
\label{3.6}
\end{equation}
The conditions (\ref{2.7}) reduce to
\begin{equation}
M\ll m_{P}.
\label{3.7}
\end{equation}

Let
\begin{equation}
{\rm I\:ii}.\qquad \qquad \qquad \alpha=1.
\label{3.8}
\end{equation}
Then
\begin{equation}
x=2.63,
\label{3.9}
\end{equation}
so that
\begin{equation}
r_{0}=2.6\,a_{0},\qquad a_{0}=a_{1},\qquad a=3.6\,a_{1}.
\label{3.10}
\end{equation}
The conditions (\ref{2.7}) reduce to
\begin{equation}
M<m_{P}.
\label{3.11}
\end{equation}

Let
\begin{equation}
{\rm I\:iii}.\qquad \qquad \qquad \alpha=8.
\label{3.12}
\end{equation}
Then
\begin{equation}
x=1,
\label{3.13}
\end{equation}
so that
\begin{equation}
r_{0}=a_{0}=8\,a_{1},\qquad a=9\,a_{1}.
\label{3.14}
\end{equation}
The conditions (\ref{2.7}) reduce to
\begin{equation}
M\lesssim m_{P}.
\label{3.15}
\end{equation}

Piecing ${\rm I\:i,ii,iii}$ together, we obtain
\begin{equation}
{\rm I}.\qquad \qquad \qquad \alpha\lesssim 10,
\label{3.16}
\end{equation}
\begin{equation}
r_{0}\gtrsim a_{0},\qquad a_{0}\lesssim 10\,a_{1}.
\label{3.17}
\end{equation}
The conditions (\ref{2.7}) reduce to
\begin{equation}
M\lesssim m_{P}.
\label{3.18}
\end{equation}

Now let
\begin{equation}
{\rm II}.\qquad \qquad \qquad \alpha\gg 10.
\label{3.19}
\end{equation}
Then
\begin{equation}
x\ll 1,\qquad x\cong\frac{1}{\alpha^{1/4}},\qquad
\frac{r_{0}}{a_{1}}=x\alpha=\alpha^{3/4}\gg 1,
\label{3.20}
\end{equation}
so that
\begin{equation}
a_{1}\ll r_{0}\ll a_{0},\qquad a\cong a_{0}.
\label{3.21}
\end{equation}
The conditions (\ref{2.7}) reduce to
\begin{equation}
a_{0}\gg l_{P}\frac{M}{m_{P}}.
\label{3.22}
\end{equation}

\section{Gravilons}

We introduce the following terminology. A gravilon is a
gravitationally autolocalized system:
\begin{equation}
{\rm gravilon}={\bf gravi}{\rm ty}+
{\bf lo}{\rm calizatio}{\bf n}.
\label{4.1}
\end{equation}
Light gravilon:
\begin{equation}
M\lesssim m_{P},
\label{4.2}
\end{equation}
heavy gravilon:
\begin{equation}
M\gg m_{P}.
\label{4.3}
\end{equation}
Fuzzy gravilon:
\begin{equation}
r_{0}\gtrsim a_{0},
\label{4.4}
\end{equation}
quasiclassical gravilon:
\begin{equation}
r_{0}\ll a_{0}.
\label{4.5}
\end{equation}

From eqs.(\ref{3.17}),(\ref{3.18}) we obtain the following
results: a fuzzy gravilon is light, a heavy gravilon is
quasiclassical. A light gravilon may be both fuzzy and
quasiclassical. A gravilon is quasiclassical iff
eqs.(\ref{3.19}),(\ref{3.22}) hold.

For a quasiclassical gravilon, we obtain from
eqs.(\ref{3.2}),(\ref{3.20}),(\ref{1.1})
\begin{equation}
r_{0}=\left( \frac{3}{4\pi} \right)^{3/4}l_{P}^{1/4}
\left( \frac{m_{P}}{\rho} \right)^{3/4}\frac{1}{a_{0}^{3/2}},
\label{4.6}
\end{equation}
so that
\begin{equation}
r_{0}\propto\frac{1}{a_{0}^{3/2}}\qquad {\rm for}
\quad \rho={\rm const}.
\label{4.7}
\end{equation}
The condition (\ref{3.22}) reduces to
\begin{equation}
a_{0}\ll 0.5\cdot 10^{14}\cdot \frac{1}{\rho^{1/2}},\qquad
[a_{0}]={\rm cm},\;[\rho]={\rm g/cm^{3}}.
\label{4.8}
\end{equation}

For the characteristic parameter $\alpha$, eq.(\ref{3.2}), we
have
\begin{equation}
\alpha=\left( \frac{4\pi}{3} \right)^{3}\frac{1}{l_{P}}
\left( \frac{\rho}{m_{P}} \right)^{3}\!a_{0}^{10}\approx
0.05\rho^{3}\left( 10^{5}a_{0} \right)^{10},\qquad
[a_{0}]={\rm cm},\;[\rho]={\rm g/cm^{3}}.
\label{4.9}
\end{equation}
Thus for $\rho\approx 1\:{\rm g/cm^{3}\;and}\;
a_{0}\gtrsim 2\cdot 10^{-5}{\rm cm}$
\begin{equation}
\alpha\gg 10,
\label{4.10}
\end{equation}
the gravilon is quasiclassical.

Let \begin{equation}
\rho=1\:{\rm g/cm^{3}},\qquad a_{0}=1\:{\rm cm};
\label{4.11}
\end{equation}
then
\begin{equation}
\alpha\approx 5\cdot 10^{48},\qquad r_{0}\approx 10^{-12}\:
{\rm cm}.
\label{4.12}
\end{equation}

Let
\begin{equation}
a_{0}=0,\qquad M=10^{-24}\:{\rm g};
\label{4.13}
\end{equation}
then
\begin{equation}
\alpha=0,\qquad r_{0}=a_{1}\approx 10^{24}\:{\rm cm}.
\label{4.14}
\end{equation}

Thus, gravilon formation, i.e., gravitational autolocalization
is predominantly a macroscopic effect.

In conclusion, it should be pointed out that the motion
of a gravilon as a whole, i.e., the motion of the
gravitational well, is classical. In the Newtonian
approximation, the internal (quantum) degree of freedom of the
center of mass is frozen: the state of the center of mass
is fixed. Thus we should not consider transitions of the
center of mass in the well.

\section*{Acknowledgment}

I would like to thank Stefan V. Mashkevich for helpful
discussions.


\begin{thebibliography}{9}

\bibitem{1} Vladimir S. Mashkevich, {\it Indeterministic
Quantum Gravity} (gr-qc/9409010, 1994).

\bibitem{2} Vladimir S. Mashkevich, {\it Indeterministic
Quantum Gravity II. Refinements and Developments}
(gr-qc/9505034, 1995).

\bibitem{3} Vladimir S. Mashkevich, {\it Indeterministic
Quantum Gravity III. Gravidynamics versus Geometrodynamics:
Revision of the Einstein Equation} (gr-qc/9603022, 1996).

\bibitem{4} Vladimir S. Mashkevich, {\it Indeterministic
Quantum Gravity IV. The Cosmic-length Universe and the
Problem of the Missing Dark Matter} (gr-qc/9609035, 1996).

\bibitem{5} Vladimir S. Mashkevich, {\it Indeterministic
Quantum Gravity V. Dynamics and Arrow of Time}
(gr-qc/9609046, 1996).

\bibitem{6} Vladimir S. Mashkevich, {\it Indeterministic
Quantum Gravity and Cosmology VI. Predynamical Geometry
of Spacetime Manifold, Supplementary Conditions for
Metric, and CPT} (gr-qc/9704033, 1997).

\bibitem{7} Vladimir S. Mashkevich, {\it Indeterministic
Quantum Gravity and Cosmology VII. Dynamical Passage
through Singularities: Black Hole and Naked Singularity,
Big Crunch and Big Bang} (gr-qc/9704038, 1997).

\bibitem{8} Murrey Gell-Mann and James Hartle,
{\it Strong Decoherence} (gr-qc/9509054, 1995).

\bibitem{9} Roger Penrose, The Emperor's New Mind
(Vintage, 1990).

\end{thebibliography}
\end{document}